\documentclass[sigconf]{acmart}

\AtBeginDocument{%
  \providecommand\BibTeX{{%
    \normalfont B\kern-0.5em{\scshape i\kern-0.25em b}\kern-0.8em\TeX}}}

\copyrightyear{2022}
\acmYear{2022}
\setcopyright{rightsretained}
\acmConference[WWW '22] {Proceedings of the ACM Web Conference 2022}{April 25--29, 2022}{Virtual Event, Lyon, France.}
\acmBooktitle{Proceedings of the ACM Web Conference 2022 (WWW '22), April 25--29, 2022, Virtual Event, Lyon, France}
\acmPrice{15.00}
\acmISBN{978-1-4503-9096-5/22/04}
\acmDOI{10.1145/3485447.3512285}

\usepackage{times}
\usepackage{latexsym}
\usepackage{amsfonts,amsmath}
\usepackage{url}
\usepackage{paralist}
\usepackage{amsmath}
\usepackage{multirow}
\usepackage{multicol}
\usepackage{caption}
\usepackage{enumerate}
\usepackage{graphicx}
\usepackage{enumitem}
\usepackage{bm}
\usepackage{url}
\usepackage[normalem]{ulem}
\usepackage{subfigure}
\usepackage{mathrsfs}

\newcommand{\argmax}{\mathop{\mathrm{argmax}}}
\newcommand{\softmax}{\mathop{\mathrm{softmax}}}
\newcommand{\Multi}{\mathop{\mathrm{Multi}}}

\acmSubmissionID{813}

\begin{document}

\title{Successful New-entry Prediction for Multi-Party Online Conversations via Latent Topics and Discourse Modeling}


\author{Lingzhi Wang}
\email{lzwang@se.cuhk.edu.hk}
\affiliation{
\institution{The Chinese University of Hong Kong}
\city{Hong Kong}
  \country{China}
}

\author{Jing Li}
\email{jing-amelia.li@polyu.edu.hk}
\affiliation{%
  \institution{The Hong Kong Polytechnic University}
  \city{Hong Kong}
  \country{China}
}

\author{Xingshan Zeng}
\email{zxshamson@gmail.com}

\affiliation{
\institution{Huawei Noah’s Ark Lab}
\city{Hong Kong}
  \country{China}
}

\author{Kam-Fai Wong}
\email{kfwong@se.cuhk.edu.hk}
\affiliation{
\institution{The Chinese University of Hong Kong}
\city{Hong Kong}
  \country{China}
}
\renewcommand{\shortauthors}{Trovato and Tobin, et al.}
\renewcommand{\shorttitle}{Successful New-entry Prediction}

\begin{abstract}
With the increasing popularity of social media, online interpersonal communication now plays an essential role in people's everyday information exchange. Whether and how a newcomer can better engage in the community has attracted great interest due to its application in many scenarios. 
Although some prior works that explore early socialization have obtained salient achievements, they are focusing on sociological surveys based on the small group. 
To help individuals get through the early socialization period and engage well in online conversations, 
we study a novel task to foresee whether a newcomer's message will be responded to by other participants in a multi-party conversation (henceforth \textbf{Successful New-entry Prediction}). The task would be an important part of the research in online assistants and social media. 
To further investigate the key factors indicating such engagement success, 
we employ an unsupervised neural network, Variational Auto-Encoder (\textbf{VAE}), to examine the topic content and discourse behavior from newcomer's chatting history and conversation's ongoing context. 
Furthermore, two large-scale datasets, from Reddit and Twitter, are collected to support further research on new-entries. 
Extensive experiments on both Twitter and Reddit datasets show that our model significantly outperforms all the baselines and popular neural models. 
Additional explainable and visual analyses on new-entry behavior shed light on how to better join in others' discussions. 
\end{abstract}






\maketitle

\section{Introduction}
Online conversations are a crucial part of our daily communication ---  many people now turn to social media to share ideas and exchange information, especially when facing the lockdown caused by an epidemic outbreak (such as the \textit{COVID-19}).  Entrance to ongoing conversations (both online and offline \cite{ahuja2003socialization}) requires a socialization process where newcomers seek to gain feedback, and the early socialization experiences have a long-term impact for newcomers \cite{burke2010membership}.
In our everyday life, one should engage in a wide variety of conversations, ranging from online meetings advancing project collaborations to chitchats forming personal ideologies. However, not everyone is good at socializing \cite{hsueh2015needs} and online newcomers face special difficulties as a result of the diffuse, decentralized, and anonymous text-based interactions \cite{burke2010membership,arguello2006talk}. It is rather challenging for a newcomer to engage in an online multi-party conversation. 

In light of these concerns, there exists a pressing need to develop a conversation management toolkit to predict the conversation's future trajectory and advance the interpersonal communication quality~\cite{jiao2018find,zeng-etal-2019-joint}.
Therefore, here we focus on a novel task to predict \textbf{successful new-entries} --- whether a newcomer's message will be replied by others and concretely contribute to the conversation's continuity. This task is inspired by some previous works, such as newcomers' socialization \cite{farzan2012socializing, burke2010membership,galvin2001doing}, social expression \cite{chua2013beyond} and response prediction \cite{artzi2012predicting,backstrom2013characterizing}, etc.  
The reasons why we employ ``receiving a reply'' to represent successful engagement can be summarized as twofold: (1) \citet{pethe2019trumpiest} shows that replies, likes (which are used to show appreciation for a post \cite{o2018rate}), retweets are highly correlated. (2) We conduct an interesting human evaluation (see Table~\ref{tab:human_eval}), and the results show that posts that received replies are more successful than posts with silent responses in four indicators. 
Hence we decide to employ such a definition to simplify the task and relieve the burden of data collection and construction.
Overall, this task can help newcomers avoid killing the conversations \cite{jiao2018find} and decrease the risk of withdrawal \cite{arguello2006talk} because the silence outcomes can be interpreted as rudeness or unfriendliness \cite{cramton2002attribution} and decline the newcomers' future participation \cite{baym1993interpreting,arguello2006talk}. More importantly, our research will potentially benefit the development of chatbots or online sales bots to understand when, what and how to say in a multi-party conversation since they should speak as few times as possible but keep the conversation going.

\begin{figure}[t]
\centering
\includegraphics[width=0.45\textwidth]{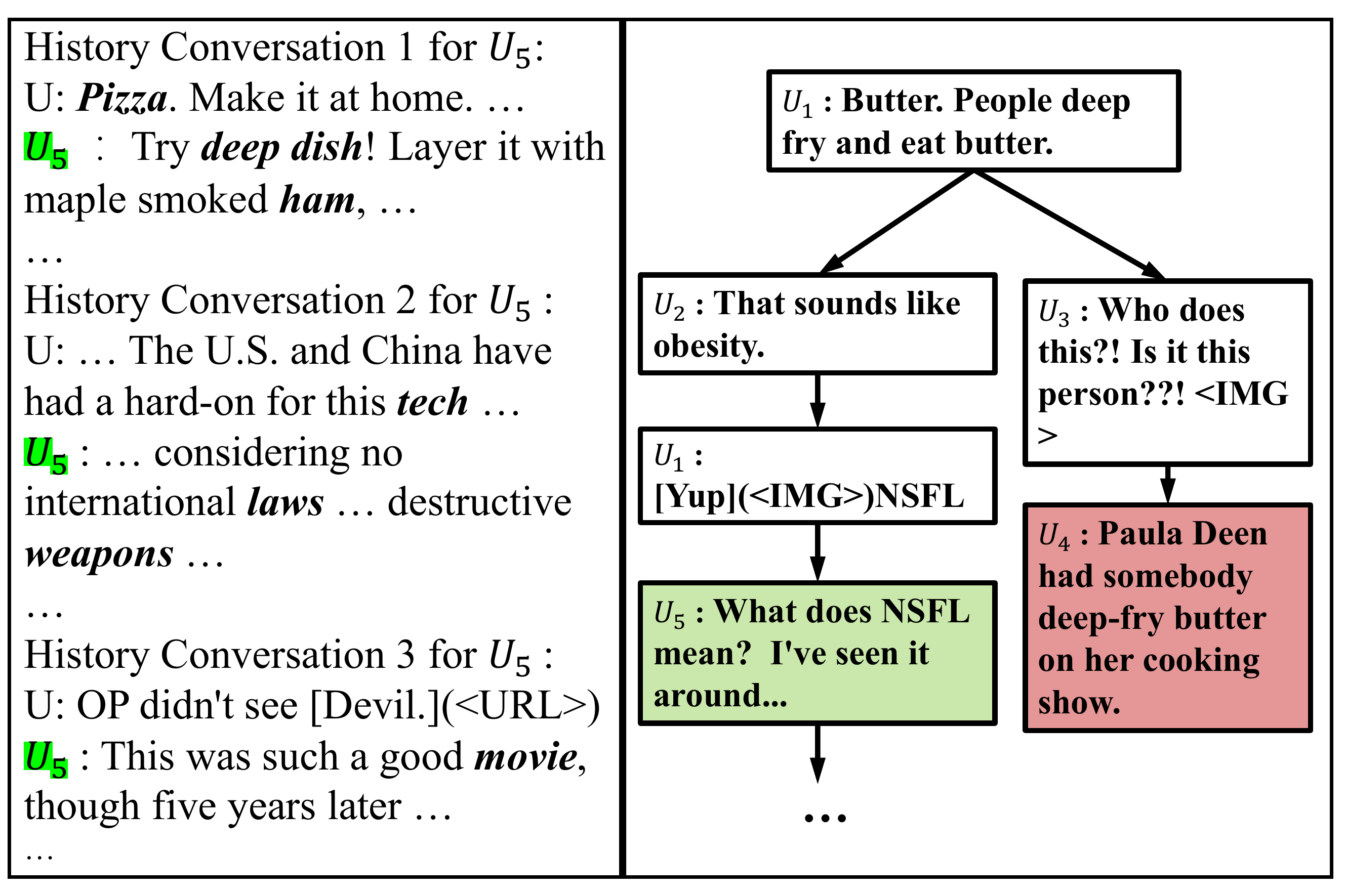}
\caption{\label{fig:intro_case}A Reddit conversation on the left part ($U_i$: the $i$-th user).
$U_3$ made a successful engagement (i.e. receiving a reply, we omit here to save space). The right column shows $U_3$'s chatting history, where the topic words are in \textit{\textbf{bold and italic}}. }
\end{figure}

To solve the successful new-entry prediction task, we propose a novel framework consisting of two parts: topic and discourse modeling (TDM) and successful new-entry prediction (SNP). 
We examine both the conversation's ongoing contexts (henceforth \textbf{conversation contexts}) and the newcomer's chatting history (\textbf{user history}) and hypothesize that both the chatting topics (what are said) and discourse behavior (how they are said) will affect other participants' attitude to the fresh blood. 
To elaborate the motivation of this hypothesis, Figure \ref{fig:intro_case} shows an example of a Reddit conversation snippet, where both $U_4$ and $U_5$ are newcomers. It is observed that $U_5$ posted a question about a new point ``NSFL'' (not safe for life) and hence drew future participants' attention and feedback, while $U_4$ made a statement via echoing ``\textit{deep-fry butter}'' concerned before and didn't receive any responses. 
To explore the factors that affect such successful and failed new entries, we capture and distinguish topic and discourse factors with an unsupervised neural module based on variational auto-encoders (VAE) \cite{kingma2013auto}. Salient words reflecting topic content (such as ``pizza'' and ``ham'' in Figure \ref{fig:intro_case}) derived from history conversations and discourse behavior (e.g., ``what'' and ``?'' as the question indicators) representing current trajectory are identified, serving as part of inputs for prediction module (i.e. the SNP module). Our SNP module contains a hierarchical two-layer Bi-GRU \cite{cho2014learning} to encode the conversation content for final prediction.
To the best of our knowledge, we are the first to study the future trajectory of newcomers-engaged conversations and how topic and discourse factors influence their engagement success (or failure). 

To summarize, the contributions of this work are as follows:
\begin{itemize}[leftmargin=*,topsep=0pt,itemsep=0pt,parsep=0pt]
    \item We first formulate the task of successful new-entry prediction and contribute two large-scale datasets, Twitter and Reddit. The SNP task can benefit the development of online assistants and early socialization strategies. 
    \item We propose a novel framework combining unsupervised and supervised neural networks. VAE and RNN-based modules are incorporated for the personalized user engagement prediction via learning latent topic and discourse representation. 
    \item Experimental results on both Twitter and Reddit show that the proposed model significantly outperforms the baselines. For example, we achieve 34.6 F1 on Reddit compared with 32.5 achieved by a BERT-based method \cite{devlin2018bert}. 
    \item Extensive analytical experiments are conducted to show the effectiveness of the modules. We probe into the learned topics and discourse and make the results explainable. Some successful entry strategies for early socialization are given based on the analysis on top of the differences between successful and failed new-entries. 
\end{itemize}




\section{Related Work}
Our work is in line with newcomer socialization, response prediction, and conversation modeling on social media platforms. 

\subsection{Newcomer Socialization}
Newcomer socialization \cite{farzan2012socializing, burke2010membership,galvin2001doing}, which analyzes the process that newcomers make the transition from being organizational outsiders to being insiders \cite{bauer2007newcomer},
is an essential research for supporting, socializing and integrating members to virtual environments \cite{horrocks2002semantic}. A range of user behaviors have been investigated, including information process \cite{ahuja2003socialization},
socialization \cite{farzan2012socializing}, and social expression \cite{chua2013beyond}. 

Most of the previous researches are based on sociological surveys in a small group while our work is based on large-scale datasets. Despite focusing on early socialization, our work also benefits the development of online assistants, e.g., chatbots. 

\begin{figure*}[t]
\centering
\includegraphics[width=0.96\textwidth]{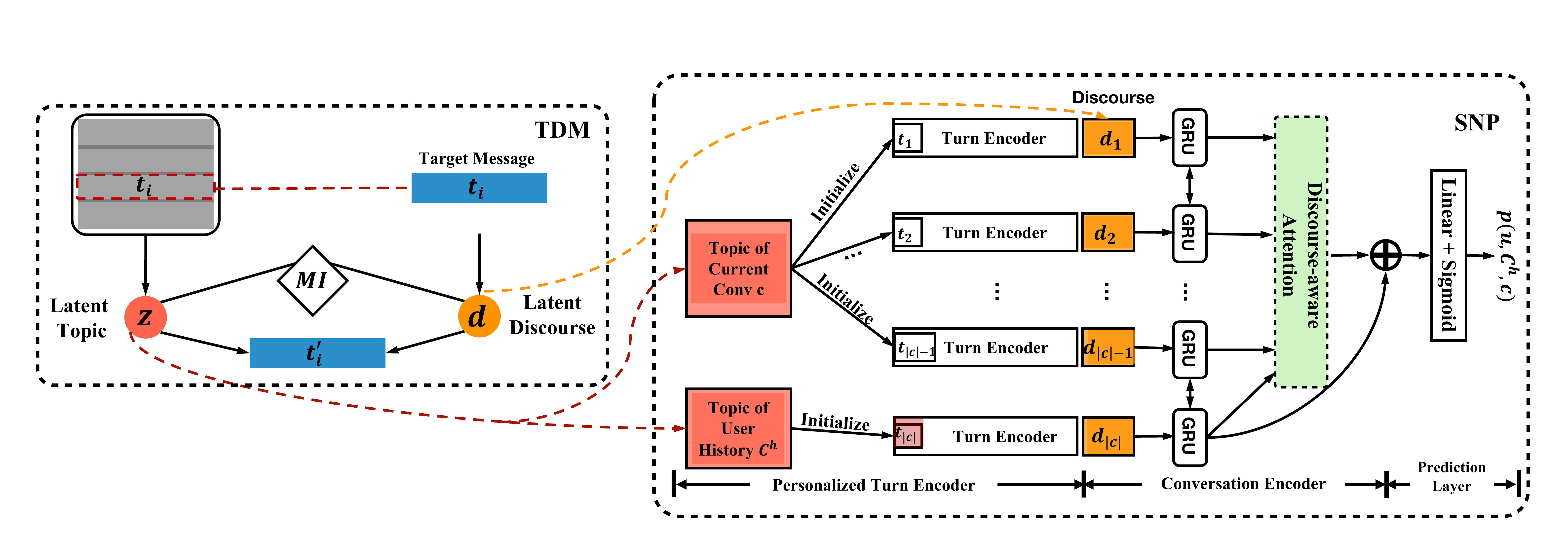}
\vskip -1.5em
\caption{\label{fig:sketch}Our generic framework for successful new-entry prediction. It contains two modules: Topic and Discourse Modeling (\textbf{TDM}) and Successful New-entry Prediction (\textbf{SNP}). SNP consists of three parts: turn encoder, conversation encoder and prediction layer.}
\end{figure*}

\subsection{Response Prediction}
Research on response prediction aims to predict whether a given online content will receive desired responses by analyzing the propagation patterns of social media content, such as the prediction of
user responses \cite{artzi2012predicting,  10.1145/2872427.2882985}, thread-ending turns  \cite{jiao2018find}, and re-entry behavior \cite{zeng-etal-2019-joint,wang-etal-2021-entry-prediction}, etc. 
Some of them tackle post-level responses \cite{rowe2011predicting, rowe2014mining, artzi2012predicting, 10.1145/2872427.2882985}, which focus on modeling single post content; and others concern users' future behavior prediction enhanced with context modeling \cite{jiao2018find, zeng-etal-2019-joint,DBLP:conf/www/Zeng0HGLK20,10.1145/3473970}, to achieve conversational level trajectory prediction. For the methodology, earlier works depend on the extraction of handcrafted features related to their objectives, e.g., social and sentiment features~\cite{artzi2012predicting}, arrival patterns and timing features~\cite{backstrom2013characterizing}, group information~\cite{budak2013participation}, etc. More recent achievements obtained from probabilistic graphical methods~\cite{bi2016modeling}, neural models like RNN-based models~\cite{jiao2018find,zeng-etal-2019-joint} and graph nerual networks~\cite{zeng-etal-2019-neural}, due to their better ability in leveraging abundant context information.

However, none of them considers newcomers in a multi-party conversation and how to help them better engage in, which is extensively investigated in this work.

\subsection{Conversation Modeling}
We are also inspired by the existing methods for conversation modeling, which develop models to encode conversation context from rich information, such as user interactions \cite{DBLP:journals/tacl/ZayatsO18,zeng-etal-2019-neural,wang-etal-2021-quotation}, temporal orders of turns \cite{cheng2017factored,jiao2018find,DBLP:conf/acl/TanakaTA19}, latent topics and discourse~\cite{zeng-etal-2019-say,wang-etal-2020-continuity}, etc. Among them, \citet{jiao2018find} and \citet{DBLP:conf/acl/TanakaTA19} only consider temporal orders of conversation turns and use RNN-based methods to learn temporal and content information, while \citet{DBLP:conf/emnlp/WeiXM19} and \citet{zeng-etal-2019-neural} formulate multi-party conversations as tree structure and apply Graph Convolutional Networks (GCN)~\cite{DBLP:conf/iclr/KipfW17} to extract context-aware representations. \citet{ritter2010unsupervised} and \citet{zeng-etal-2019-say} explore unsupervised methods to discover discourse-level information for online conversations, while the latter further leverages neural models to jointly learn distinguished topics and discourse.

In our model, the way we learn latent topic and discourse representations is built upon the success of neural latent variable models~\cite{pmlr-v70-miao17a} for unsupervised conversation understanding~\cite{zeng-etal-2019-say}. 
Compared with the original design, we also consider user chatting history to leverage newcomers' personal interests and investigate how topics and discourse factors affect newcomers' involvement in multi-party conversations, which is beyond the capability of \citet{zeng-etal-2019-say}.

\section{Our Model}
Here we present our successful new-entry prediction framework.
Figure \ref{fig:sketch} shows our overall framework. 
In the following, we first introduce the input and output in Section \ref{ssec:model:input_output}. Then in Section \ref{ssec:model:TDM}, we discuss how the TDM module works. The SNP module will be later described in Section \ref{ssec:model:NEP}.
The learning objective of the entire framework will be given at last in Section \ref{ssec:model:loss}.

\subsection{Input and Output}\label{ssec:model:input_output}
The input for our model can be divided into two parts: the observed conversation $c$ and the history conversation set $C^h = \{c^h_1, c^h_2, ..., c^h_k\}$ of the newcomer $u$, where ${k}$ is the number of history conversations obtained from training set. 
The conversation $c$ is formalized as a sequence of turns (e.g., posts or tweets) $\{t_1, t_2, ..., t_{|c|}\}$, and the $|c|^{th}$ turn is posted by the newcomer $u$ (we predict whether $u$ can get others' response afterwards). The conversations in user's history conversation set $C^h$ are organized similarly into the sequences of turns.
For output, we yield a Bernoulli distribution $p(u, C^h, c)$ to indicate the estimated likelihood of whether $u$ gets responses from other participants (successful new-entries).

\subsection{Topic and Discourse Modeling (TDM)}\label{ssec:model:TDM}
Inspired by \citet{zeng-etal-2019-say}, we learn distributional word clusters that reflect the latent topic $\bm{z}$ in conversation $c$, and discourse behaviors $\bm{d} = \langle \bm{d}_1, \bm{d}_2, ..., \bm{d}_{|c|} \rangle$ for each turn in $c$. At corpus level, we assume there are $K$ topics, each is represented by a word distribution $\phi^T_k (k = 1, 2, ..., K )$. It also contains $D$ discourse behaviors represented with word distributions $\phi^D_d (d = 1, 2, ..., D)$. 

To learn the topic and discourse representations, each turn $t_i$ in one conversation, referred to as a \textbf{target message}, is fed into the TDM sequentially in the form of bag-of-words (BoW). 
For factor modeling, TDM employs the extended framework of variational auto-encoders (VAE) \cite{kingma2013auto} to resemble the data generative process via two steps \cite{pmlr-v70-miao17a, srivastava2017autoencoding}. First, given the target turn $t_i$ and its conversation context $c$ (i.e. the other turns in the same conversation), TDM converts them into two latent variables: topic variable $\bm{z}$ and discourse variable $\bm{d}_i$. Then, we reconstruct the target turn $t_i$ with the intermediate representations captured by $\bm{z}$ and $\bm{d}_i$. In the following, we first describe the encoder followed by the decoder in detail.

\noindent \textbf{Encode step.} We learn the parameter $\mu$, $\sigma$ and $\pi$ from the input $\bm{c}_{bow}$ and $\bm{t}_{i_{bow}}$ (the BoW form of the conversation $c$ and target turn $t_i$) following the formula below:
\begin{equation}
    \begin{aligned}
        &{\bf \mu}=f_{\mu}(f_e({\bm{c}_{bow}})) , \,    log\,{\bf \sigma}=f_{\sigma}(f_e({\bm{c}_{bow}}))\\
        &{\bf \pi} = \softmax(f_{\pi}(\bm{t}_{i_{bow}}))
    \end{aligned}
\end{equation}
\noindent where $f_*(\cdot)$ is neural perceptrons performing linear transformations activated with an ReLU function~\cite{Nair:2010:RLU:3104322.3104425}.

\noindent \textbf{Decode step.} In general, the decoder learns to reconstruct the words in target message $t_i$. The following is the procedure: 
\vskip 0.5 em
\begin{itemize}
    \item Draw latent topic ${\bf z}\sim \mathcal{N}({\bf \mu}, {\bf \sigma}^2)$.
    \item  Topic mixture $\bm \theta=\softmax(f_{\theta}({\bf z}))$.
    \item  Draw the latent discourse $\bm d \sim \Multi(\pi)$.
    \item For the $n$-th word in the conversation: 
    \begin{compactitem}
        \item $\beta_n = \softmax(f_{\phi^{T}}(\bm \theta)+f_{\phi^{D}}(\bm d))$
        \item Draw the word $w_n\sim \Multi(\beta_n)$.
    \end{compactitem}
\end{itemize}
\vskip 0.5 em
\noindent In particular, the weight matrix of $f_{\phi^T}(\cdot)$ (after the softmax normalization) is considered as the topic-word distribution $\phi^T$. We can also get the discourse-word distribution $\phi^D$ in a similar way.


We use TDM to encode both contexts of the target conversation and chatting history.
For the target conversation $c$, we model and denote the topic variable as $\bm{e}_c$, and the discourse behaviors of its turns as $\bm{d} = \langle\bm{d}_1, ..., \bm{d}_{|c|}\rangle$. 
For the chatting history conversation set $C^h$ of the newcomer $u$, we learn the topic variables for all the conversations. Then they are averaged as $u$'s representation, denoted as $\bm{e}_u$. This can be regarded as a kind of user embedding, which reflects their preferences and interests learned from the user history.

\subsection{Successful New-entry Prediction (SNP)}\label{ssec:model:NEP}
This section describes how we encode conversation and predict successful new-entry via leveraging topic and discourse variables learned by TDM (described in Section \ref{ssec:model:TDM}).  
It mainly contains three parts: personalized turn encoder, discourse-aware conversation encoder, and the final prediction layer. 

\paragraph{Personalized Turn Encoder.}
For each turn $t_i$ in conversation $c$, we learn turn representations that are aware of personalized topic vector $\bm{e}_u$.
To that end, we first feed each word $w_j$ in $t_i$ into an embedding layer and get the word representation $\bm{r}_{ij}$. Then a bidirectional gated recurrent unit (Bi-GRU) \cite{cho2014learning} is used to encode the word vector sequence of turn $t_i$, denoted as $\langle \bm{r}_{i1}, \bm{r}_{i2}, ..., \bm{r}_{i,n_i}\rangle$. $n_i$ is the number of words in $t_i$.

We divide the observed conversation turns into context turns (turns before the last turn) and query turn (last turn, posted by newcomer $u$). 
For query turn, we use $u$'s topic representation $\bm{e}_u$ (produced by TDM module in Section \ref{ssec:model:TDM}) to initialize the aforementioned Bi-GRU. 
For the context turns, the topic representation $\bm{e}_c$ for conversation $c$ is similarly utilized for initialization. 
Concretely, the initial states for both directions are $\overrightarrow{\bm{h}_{i,0}} = \overleftarrow{\bm{h}_{i,n_i}} = \bm{W}^{
P}\bm{e} + b^{P}$, where $\bm{e}$ is $\bm{e}_u$ or $\bm{e}_c$.
For all turns, the hidden states of Bi-GRU are defined as:
\begin{equation}\small
\overrightarrow{\bm{h}_{ij}} = f_{GRU}(\bm{r}_{ij}, \bm{h}_{i,j-1}), \,\overleftarrow{\bm{h}_{ij}} = f_{GRU}(\bm{r}_{ij}, \bm{h}_{i,j+1})
\end{equation}
The representation of turn $t_i$ is the concatenation of last hidden states of both directions of Bi-GRU: $\bm{h}_i = [\overrightarrow{\bm{h}_{i,n_i}}; \overleftarrow{\bm{h}_{i,0}}]$. Finally, we get the turn-level representations of conversation $c$: $\langle\bm{h}_1, \bm{h}_2, .., \bm{h}_{|c|}\rangle$.

\paragraph{Discourse-aware Conversation Encoder.}
We incorporate latent discourse behavior to model the turn interactions in the conversations, which allows better understanding of how users interact with each other in the discourse. 
We first concatenate the turn-level representations with the discourse variable $\bm{d}$ learned in TDM, as the input for the second Bi-GRU layer. This Bi-GRU layer is used to model conversation structure and defined as:
\begin{equation}\small
\overrightarrow{\bm{h}_{j}^d} = f_{GRU}(\bm{s}_{j}, \bm{h}_{j-1}^d), \,\overleftarrow{\bm{h}_{j}^d} = f_{GRU}(\bm{s}_{j}, \bm{h}_{j+1})
\end{equation}
where $\bm{s}_j = [\bm{h}_j;\bm{d}_j]$ and the representation of each turn after GRU is $\bm{h}_j^d = [\overrightarrow{\bm{h}_{j}^d}; \overleftarrow{\bm{h}_{j}^d}]$. 

Then, we design an attention mechanism (henceforth discourse-aware attention) to identify discourse behavior in contexts that contribute more to signal successful engagement.  
Our intuition is that different discourse behaviors represent different functions, and therefore should be distinguished in the weights to make predictions.
For example, a turn raising a question might be more important than a simple agreement response. Therefore, we assign different attention weights to the turns, based on their discourse behaviors:
\begin{equation} \small \label{eq:att}
    a_j = f_d(\argmax(\bm{d}_j))
\end{equation}
where $\argmax(\bm{d}_j)$ means the learned discourse behavior to turn $j$, and $f_d(\cdot)$ maps the discourse behaviors to different weight values. 

Finally, to produce the whole conversation representation, we concatenate the hidden state of $|c|^{th}$ turn (i.e. the query turn) $\bm{h}^d_{|c|}$ and the weighted sum of all conversation turns:
\begin{equation} \small
    \bm{h}^c = [\bm{h}^d_{|c|}; \sum_j \softmax(a_j) \bm{h}^d_j]
\end{equation}

\paragraph{Prediction Layer.}
For prediction, we employ a linear projection function, activated by a sigmoid activation function, to predict how likely the newcomer can successfully chip in the conversation:
\begin{equation} \small
    p(u, C^h, c) = \sigma(\bm{w}^T\bm{h}^c+b)
\end{equation}
where $\bm{w}^T$ and $b$ are trainable, and $\sigma(\cdot)$ is the sigmoid activation function.

\subsection{Learning Objective}\label{ssec:model:loss}
For parameter learning in our model, we design the objective function as follows, to allow the joint learning of TDM and SNP modules:
\begin{equation}\small
\mathcal{L} = \mathcal{L}_{TDM} + \mathcal{L}_{SNP}
\end{equation}
\paragraph{Objective Function of TDM}
Following \citet{zeng-etal-2019-say}, $\mathcal{L}_{TDM}$ is defined as follows:
\begin{equation}\small
\mathcal{L}_{TDM} = \mathcal{L}_{z} + \mathcal{L}_{d} + \mathcal{L}_{t} -\lambda\mathcal{L}_{MI} 
\end{equation}
where $\mathcal{L}_{z}$ and $\mathcal{L}_{d}$ are objectives about learning topics and discourse, $\mathcal{L}_{t}$ is the loss for target message reconstruction, and $\mathcal{L}_{MI}$ ensures that topics and discourse learn differently.

To learn the latent topics and discourse, TDM employs the variational inference \cite{blei2017variational} to approximate posterior distribution over the latent topic $\bm{z}$ and the latent discourse $\bm{d}$ given all the training data. $\mathcal{L}_{z}$ and $\mathcal{L}_{d}$ is defined as follows:
\begin{equation}\small
\mathcal{L}_{z} = \mathbb{E}_{q(\bm{z}|c)}[p( c\,|\,\bm{z})] -  D_{KL}(q(\bm{z}\,|\,c)||p(\bm{z}))
\end{equation}
\begin{equation}\small
\mathcal{L}_{d} = \mathbb{E}_{q(\bm{d}|t)}[p( t\,|\,\bm{d})] -  D_{KL}(q(\bm{d}\,|\,t)||p(\bf{d}))
\end{equation}
where $q(\bm{z}|c)$ and $q(\bm{d}|t)$ are approximated posterior probabilities describing how the latent topic $\bm{z}$ and the latent discourse $\bm{d}$ are generated from the conversations and message turns. $p(c\,|\,\bm{z})$ and $p(t\,|\,\bm{d})$ represent the corpus likelihoods conditioned on the latent variables. $p(\bm{z})$ follows the standard normal prior $\mathcal{N}(\bf 0,\bf I)$ and $p(\bm{d})$ is the uniform distribution $Unif(0,1)$. $D_{KL}$ refers to the Kullback-Leibler divergence that ensures the approximated posteriors to be close to the true ones.

For $\mathcal{L}_{t}$ that ensures the learned latent topics and discourse can reconstruct target turn $t$, it is defined as below:
\begin{equation}\small
\mathcal{L}_{t} = \mathbb{E}_{q(\bm{z}|c)q(\bm{d}|t)}[log\,p(t\,|\,\bm{z},\bm{d})]
\end{equation}
We also leverage on $\mathcal{L}_{MI}$ to guide the model to separate word distributions that represent topics and discourse. $\mathcal{L}_{MI}$ is defined as:
\begin{equation}\small
\mathcal{L}_{MI} = \mathbb{E}_{q(\bm{z})}D_{KL}(p(\bm{d}\,|\,\bm{z})||p(\bm{d}))
\end{equation}
\paragraph{Objective Function of SNP}
The objective function of SNP is designed to be binary cross-entropy loss as following:
\begin{equation}\small
\mathcal{L}_{SNP} = -\sum_i{\mu\,y_i log(\hat{y_i})+(1-y_i)log(1-\hat{y_i})}
\end{equation}
where $\hat{y_i}$ denotes the probability estimated from $p(u, C^h, c)$ for the i-th instance, and $y_i$ is the corresponding binary ground truth label (1 for successful entries and 0 for the opposite).
To take the potential data imbalance into account, we also adopt a trade-off weight $\mu$ to give more weight to the minority class. $\mu$ is set based on the proportion of positive and negative instances in the training set. 

\section{Experimental Setup} \label{sec:setup}

\subsection{Datasets.}
We construct two new conversation datasets from \textbf{Twitter} and \textbf{Reddit}. The raw data for the \textbf{Twitter} dataset is released by \citet{zeng2018microblog}, containing Twitter conversations formed based on the TREC 2011 Microblog Track\footnote{\url{https://trec.nist.gov/data/tweets/}}. The raw data for the \textbf{Reddit} dataset contains posts and comments from Jan to May 2015, which is obtained from a publicly available Reddit corpus\footnote{\url{https://files.pushshift.io/reddit/comments/}}. For both datasets, we follow the common practice to form conversations with in-reply-to relations~\cite{zeng2018microblog,wang-etal-2019-topic-aware}, where a post or a reply (comment) is considered as a conversation turn. 

Our work focuses on the multi-party conversations with new users engaging in later.
To that end, we remove the conversations with $<4$ turns and those with $<3$ participants. 
Finally, the datasets are randomly divided into $80$\%, $10$\%, and $10$\%, for training, validation, and test.
\begin{table}[t]\large
\setlength{\tabcolsep}{3.5mm}
\newcommand{\tabincell}[2]{\begin{tabular}{@{}#1@{}}#2\end{tabular}}
\caption{
 \label{statistics-table1}Statistics of Twitter and Reddit datasets.
}
\begin{center}
\begin{tabular}{lrr}
\hline 
&  \textbf{Twitter} & \textbf{Reddit} \\
\hline
{\# of users}& 53,488 & 96,001  \\
{\# of convs} & 37,339 & 69,428  \\
{\# of conv turns} & 179,265 & 236,764 \\
{\# of successful entries} & 29,340 & 12,199 \\
{\# of failed entries} & 7,999 & 57,229 \\
\hline
{Avg turn number per conv} & 4.8 & 3.4  \\
{Avg token number per turn} & 20.5 & 20.7 \\
{Ratio of newcomer with history} & 0.59 & 0.62 \\
{Avg \# of history for newcomers} & 2.5 & 6.3\\
\hline
\end{tabular}
\end{center}
 
\end{table}
\begin{figure}[t]
\centering
\includegraphics[width=0.45\textwidth]{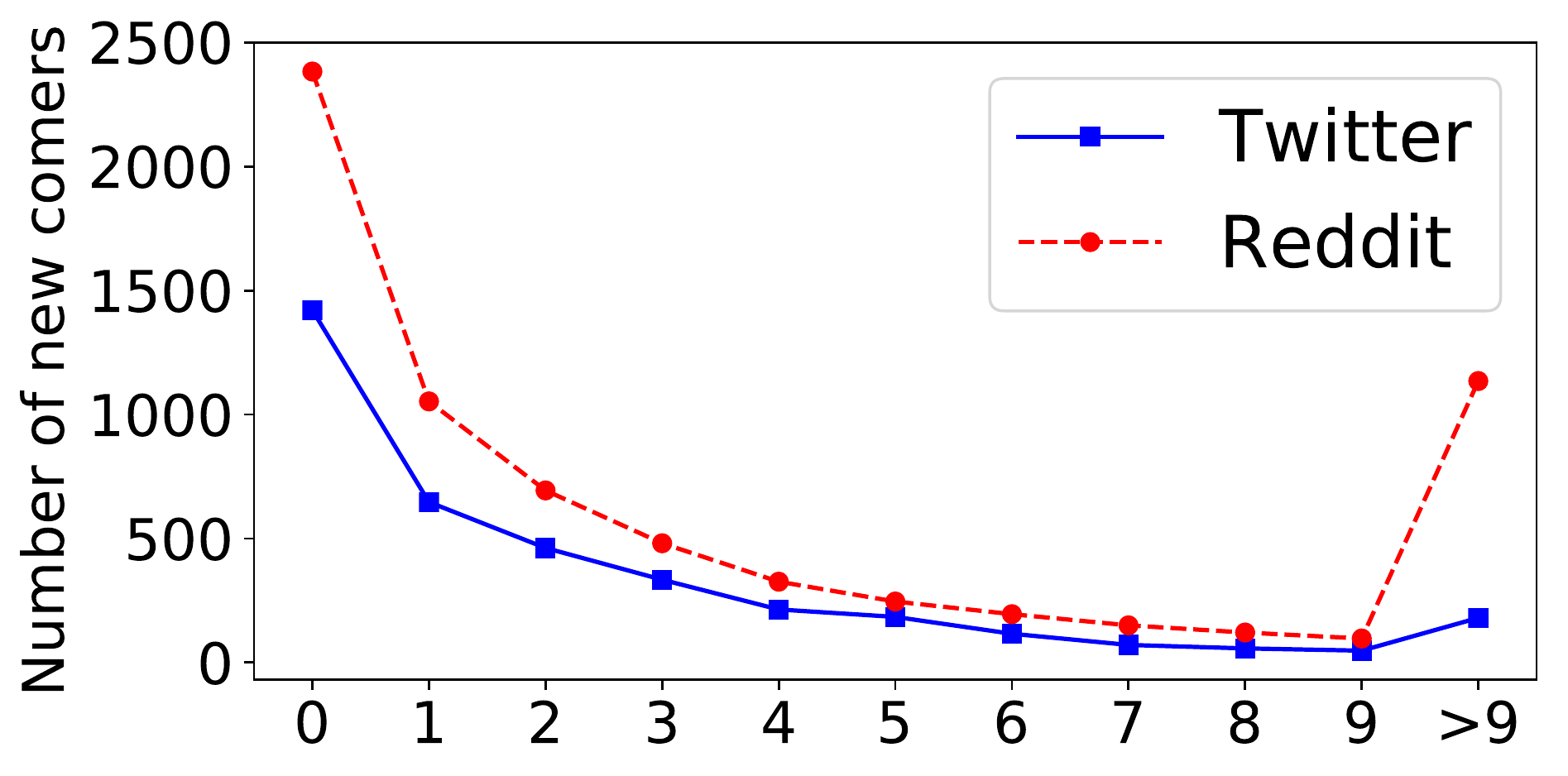}
\caption{\label{fig:data_fre}The distribution over the number of history conversations (X-axis). Y-axis: number of newcomers.}
\end{figure}

The statistics of the two datasets are shown in Table \ref{statistics-table1}.  
As can be seen, Twitter users tend to respond to newcomers while new-entries in Reddit are more likely to be failed, probably because Twitter users are more open to public discussions compared with Reddit.
We can also see that about $60$\% newcomers has user chatting history, which means that $60$\% newcomers in the test set are involved in other discussions in training data. 
We further study the newcomer's distribution for the number of history conversations in Figure \ref{fig:data_fre}.
Most of the newcomers engaged in less than $5$ conversations before.
The sparsity in user history might pose challenges to learn their interests.

\subsection{Preprocessing.}
For Twitter dataset, we applied Glove tweet preprocessing toolkit \cite{pennington2014glove}. As for Reddit, we first utilized the open-source natural language toolkit (NLTK) \cite{10.3115/1118108.1118117} for word tokenization. We then removed all the non-alphabetic tokens and replaced links with the generic tag ``URL''. For both datasets, a vocabulary was built and maintained with all the remaining tokens, including emoticons and punctuation. For the TDM module of our model, 
all stopwords were removed for topic modeling following common practice~\cite{ Blei:2003:LDA:944919.944937}.

\subsection{Parameter Setting.} 
The parameters in the TDM module are set up following \citet{zeng-etal-2019-say}. For the parameters in SNP module, we first initialize the embedding layer with 200-dimensional Glove embedding \cite{pennington2014glove}, whose Twitter version is used for Twitter dataset and Common Crawl version is applied for Reddit\footnote{\url{https://nlp.stanford.edu/projects/glove/}}. For the Bi-GRU layer, we set the size of its hidden states for each direction to 100 (200 for final output). The batch size is set to 64. In model training, we employ Adam optimizer ~\cite{kingma:adam} with initial learning rate selected among $\{1e$-$3$, $1e$-$4, 1e$-$5\}$ and early stop adoption~\cite{caruana2001overfitting}. 
Dropout strategy~\cite{Srivastava:2014:DSW:2627435.2670313} is used to alleviate overfitting. All the hyper-parameters are tuned on the validation set by grid search.

\begin{figure}[t]
\centering
\includegraphics[width=0.48\textwidth]{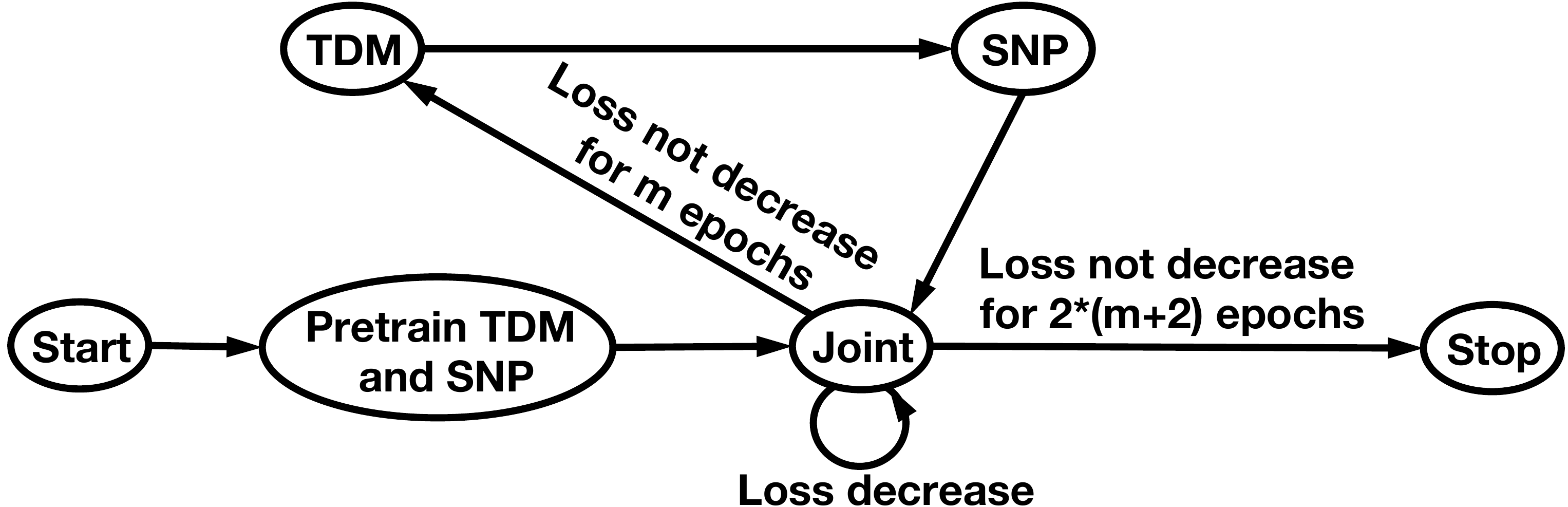}
\vskip -0.5em
\caption{\label{fig:training}State transition diagram of training process.}
\vskip -1em
\end{figure}
\subsection{Joint Train Procedure.}
How to jointly train the TDM and SNP modules is critical for the performance of our model. We design to first train the TDM module for $N_1$ epochs (e.g., $N_1 = 100$ ) and train SNP module for $N_2$ epochs (e.g., $N_2 = 5$) respectively (while the other one is fixed), as pre-training procedure. Then we jointly train the whole model alternatively. Figure \ref{fig:training} is a state transition diagram showing the process of our joint training method.

\begin{table*}[ht]\large
\setlength{\tabcolsep}{2.5mm}
\newcommand{\tabincell}[2]{\begin{tabular}{@{}#1@{}}#2\end{tabular}}
\caption{\label{tab:main} Comparison results on Twitter and Reddit datasets (in \%). Higher scores indicate better performance. The best results in each column are in \textbf{bold}. Our model gets significantly better scores than all other comparisons for all metrics ($p $$<$$0.01$, paired t-test).
}
\vskip -1em
\begin{center}
\scalebox{0.9}{
\begin{tabular}{lcccccccc}
\hline 
\multirow{2}{*}{Models} 
&\multicolumn{4}{c}{ \tabincell{c}{\textbf{Twitter}} }
&\multicolumn{4}{c}{ \tabincell{c}{\textbf{Reddit}} }

\\
\cmidrule(lr){2-5}\cmidrule(lr){6-9}
& AUC  & Accuracy & Precision & F1 & AUC &  Accuracy & Precision & F1 \\
\hline
\underline{\textbf{Simple Baselines}} & & &&& & & &\\
\textsc{Random} & 50.1 & 49.5 & 78.0 &  60.1
& 49.9 & 50.2 & 15.4 & 24.1 \\
\textsc{History} & 44.1 & 41.2 & 73.2 & 50.6
& 53.6 & 46.5 & 18.1 & 27.5 \\
\hline
\underline{\textbf{Comparisons}} & & &&& & & &\\
\textsc{SVM} & 51.5 & 56.3 & 75.7 & 74.2
& 54.3 & 50.1 & 18.9 & 29.1 \\
\textsc{BiLSTM}& 52.5 & 77.4 & 78.2 & 87.2
& 59.3 & 54.1 & 22.2 & 31.7\\
\textsc{BERT} & 70.5 & 80.2 & 80.4 &  89.0
& 63.2 & 51.1 & 21.6 & 32.5 \\
\textsc{ConverNet} & 73.6 & 79.2 & 78.9 &  88.2
& 60.6 & 55.3 & 21.6 & 31.2 \\
\textsc{JECUH}& 75.2 & 80.1 & 80.3 & 88.4
& 60.7 & 57.6 & 22.6 & 31.9\\

\hline
\textbf{Our Model} & \textbf{83.2} & \textbf{82.9} & \textbf{84.7}&\textbf{90.2}
& \textbf{64.8} & \textbf{62.7}& \textbf{24.9} & \textbf{34.6}
\\
\hline
\end{tabular}
}
\end{center}

\end{table*} 
\subsection{Comparisons and Evaluation Metrics}
\paragraph{Comparisons} We first compare with two simple baselines: \textbf{\textsc{Random}}, which randomly selects label $0$ or $1$ for prediction;  \textbf{\textsc{History}}, which predicts based on the ratio of successful entries in newcomers' history conversations (random prediction is adopted for newcomers without user history).

We further compare our model with $5$ different models. 

1) \textbf{\textsc{SVM}}: SVM-based binary classifier \cite{cortes1995support} with features (e.g., TF-IDF, topic distribution, post length, thread length, etc.) gathering from \citet{jiao2018find}, \citet{suh2010want}, and \citet{hong2011predicting}, 

2) \textbf{\textsc{BiLSTM}}: hierarchical BiLSTM used to encode conversation contexts --- a BiLSTM as turn encoder and another BiLSTM to model the turn sequence, and an MLP layer works for the predictions (like our model).

3) \textbf{\textsc{BERT}}: a pretrained BERT is adopted to learn the turn representations and another BiLSTM to model the whole conversation, an MLP layer works for the predictions (like our model).

4) \textbf{\textsc{ConverNet}}: the state-of-the-art model to predict conversation killers~\cite{jiao2018find}, where a few features are ignored (sentiment, background, etc.) because they are unavailable in our datasets. 

5) \textbf{\textsc{JECUH}}: the state-of-the-art model to predict conversation re-entries~\cite{zeng-etal-2019-joint}, where the implementation is based on their code.


\paragraph{Evaluation Metrics} To measure the performance, we adopt popular evaluation metrics for binary classification and consider area under the ROC Curve (AUC), accuracy, precision, and F1 scores. 
\section{Experimental Results}
We first compare model performance in Section \ref{ssec:result:main_result} including an ablation study. 
Then in Section \ref{ssec:result:topic_disc}, we analyze the effects of topics and discourse over successful new-entries.
Section \ref{difference} shows the differences between successful and failed cases, and verifies with a human evaluation.
At last, Section \ref{ssec:result:discussion} gives more discussions on our model output with quantitative and qualitative analyses.
\subsection{Main Comparison Results}\label{ssec:result:main_result}

Table \ref{tab:main} reports the main results on the two datasets, where our model significantly outperforms all comparison models. Here are more observations:

$\bullet$~\textit{All models perform better on Twitter than Reddit.} This suggests models' sensitivity to label imbalance, where on Reddit, we observe more sparse positive samples to learn successful new-entries compared with the negative (see Table \ref{statistics-table1}). 

$\bullet$~\textit{Successful New-entry Prediction (SNP) is challenging.} Simple baselines such as \textsc{Random} and \textsc{History} perform poorly. This indicates that SNP is challenging and impossible to be well tackled relying on simple strategies.
We also find that neural models exhibit better performance than non-neural, probably benefited from their ability to learn deep semantic features from complex online interactions, which is beyond the capability of shallow features crafted manually. 



$\bullet$~\textit{Newcomers' chatting history is useful.}
By leveraging user history, \textsc{JECUH} and our model perform better than both \textsc{BiLSTM} and ConverNet.
It might be attributed to the continuity of user interests, which allows the use of more context from user history to better understand the new-entry and how it is related to the conversation contexts.  

$\bullet$~\textit{The coupled effects of topics and discourse are helpful to SNP.} It is seen that the joint modeling of topics and discourse helps our model to obtain the best performance. 
More performance gain is observed on Reddit dataset, suggesting our ability to alleviate the overfitting caused by sparse positive samples, possibly because richer features can be learned from latent topic and discourse clusters.


\paragraph{Ablation Study.}
\begin{table}[t]\large
\setlength{\tabcolsep}{1.5mm}
\newcommand{\tabincell}[2]{\begin{tabular}{@{}#1@{}}#2\end{tabular}}
\caption{\label{tab:ablation} Comparison results with ablations(in \%). Higher scores indicate better performance. 
}
\vskip -1em
\begin{center}
\scalebox{0.9}{
\begin{tabular}{p{32mm}rrrr}
\hline 
\multirow{2}{*}{Models} 
&\multicolumn{2}{c}{ \tabincell{c}{\textbf{Twitter}} }
&\multicolumn{2}{c}{ \tabincell{c}{\textbf{Reddit}} }

\\
\cmidrule(lr){2-3}\cmidrule(lr){4-5}
& Accuracy & F1 &  Accuracy & F1 \\
\hline
\textsc{w/o Topic Init} 
& 79.5 & 88.1 & 60.5 & 32.7 \\
\textsc{w/o Disc Concat} 
& 80.8 & 88.3 & 61.4 & 33.2 \\
\textsc{w/o Disc Att} 
& 81.1 & 88.6 & 60.2 & 33.5\\
\textbf{Full Model} & \textbf{82.9} & \textbf{90.2}
 & \textbf{62.7} & \textbf{34.6}\\
\hline

\end{tabular}
}
\end{center}

\vskip -1em
\end{table}
We also examine the contributions of some components in our framework with an ablation study presented in Table \ref{tab:ablation}.
We compare our full model with its variants without using topics for turn encoder initialization (\textsc{w/o Topic Init}), without concatenating discourse as turn representation (\textsc{w/o Disc Concat}) and without discourse-aware attention (\textsc{w/o Disc Att}).
The results indicate that topic factors might contribute more than discourse (\textsc{w/o Topic Init} performs the worst), while their joint effects help exhibit the best results.
We can also find that newcomers' user history is crucial to the model, as it is used to learn their personal interests and infer new-entries' topics.



\subsection{Effects of Topic and Discourse }\label{ssec:result:topic_disc}

\begin{table}[t]\large
\setlength{\tabcolsep}{2.5mm}
\newcommand{\tabincell}[2]{\begin{tabular}{@{}#1@{}}#2\end{tabular}}
\caption{\label{tab:topic_cv} $C_v$ scores for Top 5 and 10 words of leaned topics. The values range from 0.0 to 1.0, and higher scores indicate better topic coherence.
}
\vskip -2em
\begin{center}
\scalebox{0.9}{
\begin{tabular}{lcccc}
\hline 
\multirow{2}{*}{Models} 
&\multicolumn{2}{c}{ \tabincell{c}{\textbf{Twitter}} }
&\multicolumn{2}{c}{ \tabincell{c}{\textbf{Reddit}} }

\\
\cmidrule(lr){2-3}\cmidrule(lr){4-5}
& 5 & 10 & 5 & 10 \\
\hline
\textsc{LDA} 
& 0.498 & 0.393 & 0.483 & 0.377\\
\textsc{NTM} 
& 0.499 & 0.425 & 0.492 & 0.397 \\
\textbf{Our} & \textbf{0.504} & \textbf{0.431}
 & \textbf{0.495} & \textbf{0.412}\\
\hline

\end{tabular}
}
\end{center}
\vskip -1em

\end{table}
\begin{table}[t]
\setlength{\tabcolsep}{3.5mm}
\renewcommand\arraystretch{1.2}
\newcommand{\tabincell}[2]{\begin{tabular}{@{}#1@{}}#2\end{tabular}}
\caption{\label{tab:discourse}
$5$ sample latent discourse behaviors discovered from Reddit (The top $5$ terms by likelihood are shown here). Names in the first column are our interpretation of the discourse behaviors according to the learned clusters. Discourse words indicating the behavior are highlighted in \textcolor{blue}{\textit{blue and italic}}.
}
\vskip -2em
\begin{center}
\begin{tabular}{ll}
\hline
\textbf{Discourse} & \textbf{Top 5 representative terms} \\
\hline
{Disagreement}& {\textcolor{blue}{\textit{but}} have ask  \textcolor{blue}{\textit{different}} see }  \\
\hline
{Explanation}& {\textcolor{blue}{\textit{because}} stil when of that}  \\
\hline
{Opinion}& {\textcolor{blue}{\textit{think}} my here ! \textcolor{blue}{\textit{never}} } \\
\hline
{Doubt}& {\textcolor{blue}{\textit{n’t}} always want like \textcolor{blue}{\textit{why}}} \\
\hline
{Question}& {\textcolor{blue}{\textit{?}} For ! \textcolor{blue}{\textit{where}} \textcolor{blue}{\textit{what}}} \\
\hline
\end{tabular}\large
\end{center}

\end{table}
\paragraph{Topic Analysis.}
To analyze the topics extracted from our TDM model, we adopt $C_v$ scores measured via the open-source Palmetto toolkit \footnote{https://github.com/dice-group/Palmetto} to evaluate the topic coherence. $C_v$ scores assume the top N words in a coherent topic (ranked by likelihood) tend to co-occur in the same document and have shown comparable evaluation results to human judgments \cite{roder2015exploring}. We present $C_v$ scores for top 5 and 10 words of the learned topics in Table \ref{tab:topic_cv} and compare them with LDA \cite{griffiths2004hierarchical} and NTM \cite{pmlr-v70-miao17a}. The comparison results show that our model can learn a better topic representation. 

\begin{figure}[t]
\centering
\includegraphics[width=0.35\textwidth]{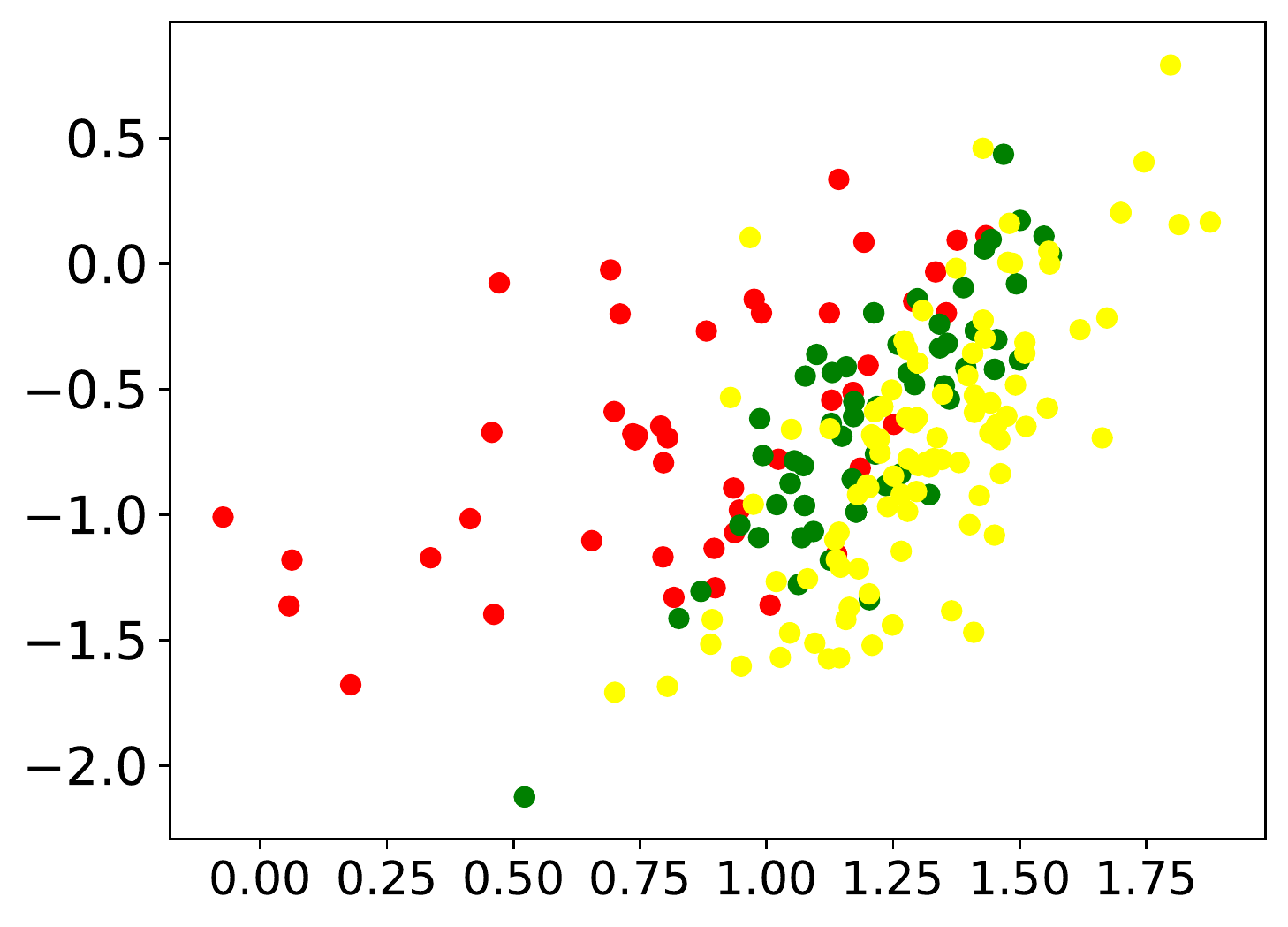}
\vskip -1em
\caption{\label{sfig:exp:2d_topic}  Topic mixture visualization (before softmax normalization) of three users' history conversations. A point refers to a conversation while different users are in distinguished colors.}
\end{figure}

To further analyze the topics learned from user history, we set the topic number $K=2$, sample three users, and visualize their history conversations' topic mixtures in Figure \ref{sfig:exp:2d_topic}.
It is seen that users tend to be drawn by discussions with similar topics, which interprets why incorporating chatting history allows better understanding of the new-entries' topics.

\paragraph{Discourse Analysis.}
\vskip -1em
Compared with latent topics, discourse behavior is harder to be understood.
So to interpret what is learned for discourse, Table \ref{tab:discourse} shows 5 example discourse behaviors from Reddit with the top 5 terms by likelihood, where meaningful discourse words are found to represent different user discourse behaviors.

\begin{figure}[t]
\centering
\subfigure[Disc distribution for newcomers]{\label{sfig:exp:disc}
\includegraphics[width=0.45\textwidth]{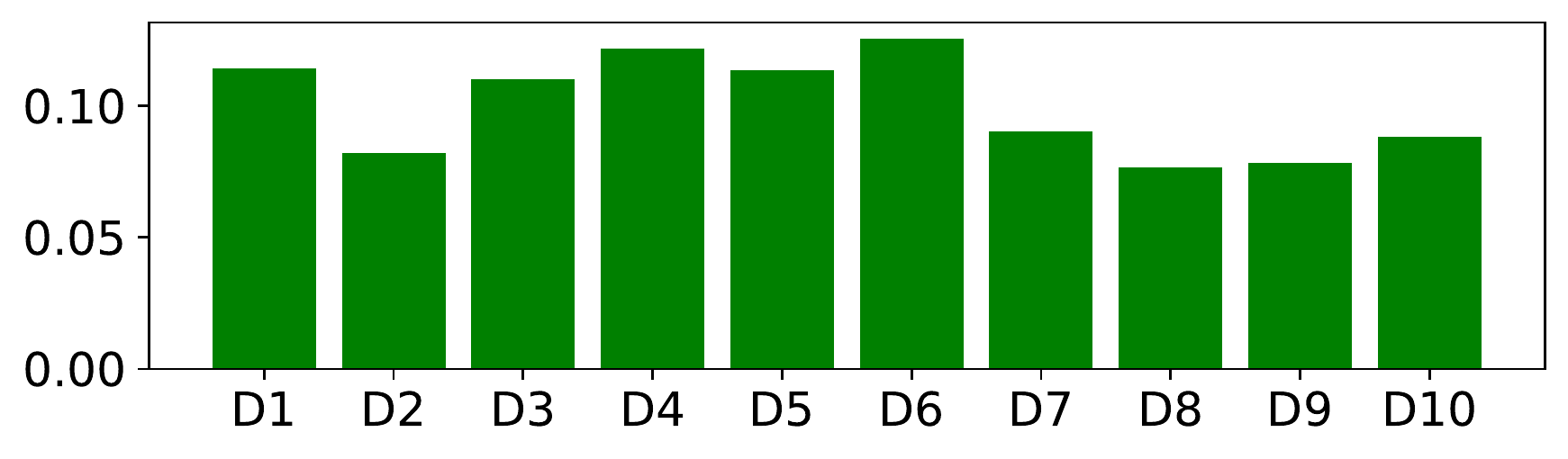}
}
\vskip -0.5em
\subfigure[Distribution for successful and failed new-entries]{\label{sfig:exp:yes_no}
\includegraphics[width=0.48\textwidth]{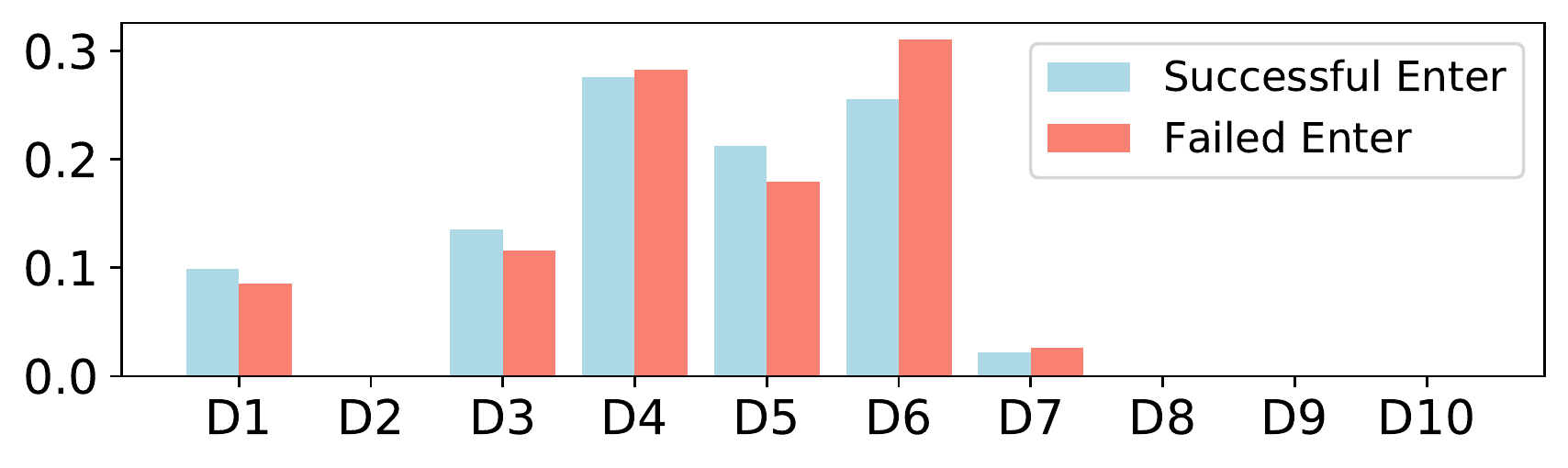}
}
\vskip -1.5em
\caption{
\label{fig:exp:distribution}
\ref{sfig:exp:disc} is the distribution over discourse behaviors used in new-entries. \ref{sfig:exp:yes_no} is the distribution of discourse behaviors for successful and failed new-entries. For both, X-axis: the 10 discourse behavior learned by our model; Y-axis: corresponding probability.
}
\end{figure}

To further examine newcomers' discourse behavior, Figure \ref{sfig:exp:disc} shows the distribution of the latent discourse learned for new-entries. 
The results show that all discourse behaviors have similar frequency to be used
, while some are relatively more popular, such as D4 and D6. 
We then compare the distribution of discourse behavior over successful and failed new-entries to Reddit conversations in Figure \ref{sfig:exp:yes_no} (the top 2 discourse predictions are considered).
As can be seen, some discourse behaviors tend to result in successful outcomes (e.g., D3 and D5), while some may easily lead to the opposite (e.g., D4 and D6).
This indicates the importance for newcomers to use the right manner so as to better chip in others' discussions.


\subsection{Successful and Failed New-entry} \label{difference}
\paragraph{Differences in Topic Similarity.}
We carry out another study to examine how topics affect new-entries. 
We measure the similarity between the newcomers' topic distribution in their chatting history and the target conversations' contexts for successful cases (SN) and the failed ones (FN). 
It turns out that the similarity scores for SN and FN are $38.7$ and $42.7$ in Twitter, while in Reddit they are $20.0$ and $22.8$ respectively. Interestingly, lower similarity results in more chances for success --- newcomers mentioning points that are discussed less in prior history (probably cutting in with a question) are more likely to be responded to by others. 
\begin{table}[t]\large
\setlength{\tabcolsep}{2.2mm}
\newcommand{\tabincell}[2]{\begin{tabular}{@{}#1@{}}#2\end{tabular}}
\caption{\label{tab:human_eval} Human evaluation results (\%). The overall inter-rater agreement achieved Krippendorff’s $\alpha$ of 0.74, which indicates reliable results~\cite{krippendorff2004content}.
}
\vskip -1.5em
\begin{center}
\scalebox{0.9}{
\begin{tabular}{p{18mm}rrrrrrrr}
\hline 
\multirow{2}{*}{Models} 
&\multicolumn{4}{c}{ \tabincell{c}{\textbf{Twitter}} }
&\multicolumn{4}{c}{ \tabincell{c}{\textbf{Reddit}} }

\\
\cmidrule(lr){2-5}\cmidrule(lr){6-9}
& OT &AQ &CL &CS & OT &AQ &CL &CS\\
\hline
\textsc{Successful} 
& 100 & 28 & 0& 54 &98&48&14&38\\
\textsc{Failed} 
& 96& 4& 0 & 6&88&20&22&20\\
\hline

\end{tabular}
}
\end{center}

\vskip -1em
\end{table}
\begin{figure}[t]\large
\centering
\subfigure[F1 with Varying History \#]{\label{sfig:exp:his_f1}
\includegraphics[width=0.32\textwidth]{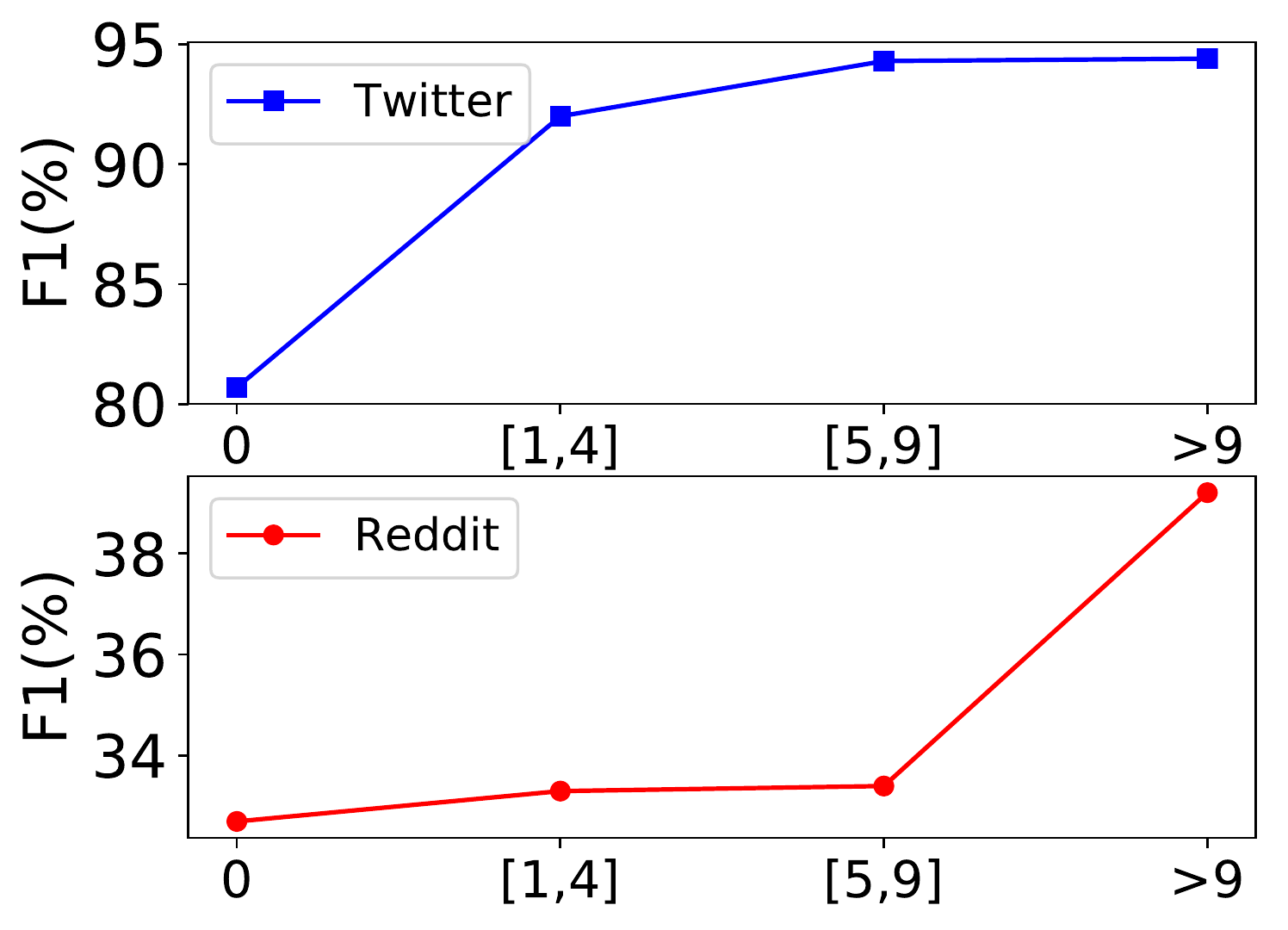}
}
\subfigure[F1 with Varying Turn \#]{\label{sfig:exp:con_len_f1}
\includegraphics[width=0.32\textwidth]{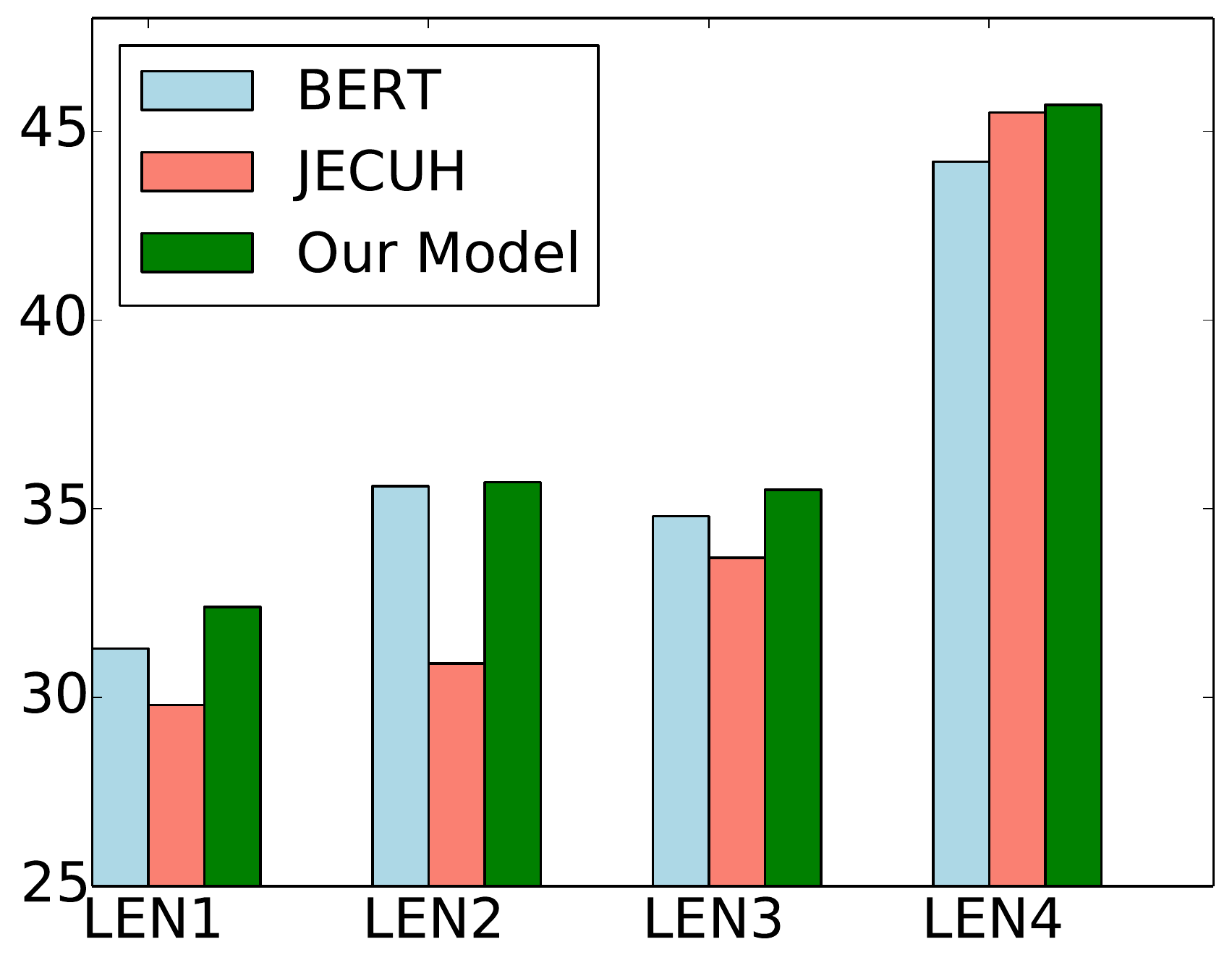}
}
\vskip -0.5em
\label{fig:hist_con_len}
\caption{
 Y-axis: F1 score. In \ref{sfig:exp:his_f1}, X-axis: user history conversation numbers. In \ref{sfig:exp:con_len_f1}, LEN$i$ in X-axis: the $i$-th quantile by turn numbers (smaller $i$, shorter length). 
}
\vskip -1em
\end{figure}

\paragraph{Human Evaluation.}
We conduct a human evaluation to explore the differences between successful and failed new-entry. Four postgraduate volunteers proficient in English are invited, and 50 conversations (25 conversations are successful cases and the others are failed) are randomly sampled from Twitter and Reddit datasets, respectively. Inspired by  \citet{arguello2006talk} and \citet{burke2010membership}, we propose four evaluation indicators, on-topics (OT), asking questions (AQ), complex language (CL) and controversial statement (CS). Volunteers give 0 (not match) or 1 (match) for each conversation according to the four indicators. From the results showed in Table \ref{tab:human_eval}, the posts of successful newcomers are more on-topic, asking questions more often, in simple language, and usually more controversial, which is consistent with \citet{arguello2006talk}. 




\subsection{Further Discussion }\label{ssec:result:discussion}


\paragraph{Effect of History Number.}
As discussed in Section \ref{ssec:result:main_result} and \ref{difference}, user history is essential to learn topic factors. We further analyze the change of prediction results over varying user history lengths. 
Test set is then divided into four subsets with the number of history conversations ($x$) involving a newcomer, where $x=0$, $1-4$, $5-9$, and $>9$. Our F1 scores over them are displayed in Figure \ref{sfig:exp:his_f1}.
Better F1 is achieved on newcomers with longer history (engaged in more conversations), as sparse history provides limited contexts to learn topics, which will further affect SNP performance.

\paragraph{Effect of Conversation Length.}
After showing how our model performs over varying sparsity degree of user history (in Figure \ref{sfig:exp:his_f1}), 
here we are interested in the model sensitivity over varying lengths of conversation contexts.
Figure \ref{sfig:exp:con_len_f1} shows the F1 scores over varying turn numbers in conversation contexts in Reddit. 
Better F1 scores are observed for conversations with more turns, as longer contexts can benefit feature learning.
Also, more performance gain is observed from our model for LEN1 (very short contexts), signaling our ability to cope with data sparsity. 


\paragraph{Qualitative Analysis.}
To provide more insights,
we use the example in Figure \ref{fig:intro_case} to conduct a qualitative analysis.
It is found that our model assigns more attention weights ($a_j$ in Eq.\ref{eq:att}) to the turn posted before $U_3$ as it concerns the topic ``NSFL'' leading to successful new-entry.
We also notice that $U_3$ has a wide range of interests (shown in user history) and open to engaging in different types of discussions. 
This might be another reason why our model predicts positive outcomes, following our previous findings (Section \ref{ssec:result:topic_disc}).

\paragraph{Advises for Newcomers.} As we showed in Section \ref{difference} that successful and failed newcomers show differences in topic similarity and four evaluation indicators, we give two suggestions as follows: 1) Contribute new and interesting information to the community, even it's controversial. 2) Use simple language and ask on-topic questions more for better communication.



\section{Conclusion}
This paper first formulate the task of successful new-entry prediction and collect two large-scale datasets, Twitter and Reddit. 
A novel model is proposed to predict successful new-entries via modeling latent topics and discourse in conversation contexts and user chatting history. This paper also explores the roles of topic and discourse played in newcomers' engagement to multi-party conversations. 
Extensive experiments have shown that the proposed model achieves significantly better performance than baselines and the model has learned meaningful topics and discourse representations, which are able to further signal how to make successful new-entries. 

\section*{Acknowledgements}
The research described in this paper is partially supported by RGC GRF \#14204118 and RGC RSFS \#3133237. Jing Li is supported by NSFC Young Scientists Fund (62006203). 

\bibliographystyle{ACM-Reference-Format}
\bibliography{anthology,custom}

\end{document}